\documentclass[twocolumn,showpacs,preprintnumbers,amsmath,amssymb]{revtex4}

\usepackage{graphicx}
\usepackage{dcolumn}
\usepackage{bm}
\usepackage{booktabs}
\usepackage{multirow}
\usepackage{amssymb}

\begin{document}


\title{Effect of Co-Fe substitutions on the room-temperature spin polarization in Co$_{3-x}$Fe$_{x}$Si Heusler-compound films}

\author{K. Tanikawa,$^{1}$ S. Oki,$^{1}$ S. Yamada,$^{1}$ K. Mibu,$^{2}$ M. Miyao,$^{1,3}$ and K. Hamaya$^{1}$\footnote{E-mail: hamaya@ed.kyushu-u.ac.jp}}

\affiliation{%
$^{1}$Department of Electronics, Kyushu University, 744 Motooka, Fukuoka 819-0395, Japan}%
\affiliation{%
$^{2}$Nagoya Institute of Technology, Nagoya, Aichi 466-8555, Japan}%
\affiliation{%
$^{3}$CREST, Japan Science and Technology Agency, Sanbancho, Tokyo 102-0075, Japan}%

\date{\today}

\begin{abstract}
Using low-temperature molecular beam epitaxy, we study substitutions of Fe atoms for Co ones in Co$_{3-x}$Fe$_{x}$Si (1.0 $\le$ $x$ $\le$ 3.0) Heusler-compound films grown on Si and Ge. Even for the low-temperature grown Heusler-compound films, the Co-Fe atomic substitution at A and C sites can be confirmed by the conversion electron M\"ossbauer spectroscopy measurements. As a result, the magnetic moment and room-temperature spin polarization estimated by nonlocal spin-valve measurements are systematically changed with the Co-Fe substitutions. This study experimentally verified that the Co-Fe substitution in Co$_{3-x}$Fe$_{x}$Si Heusler compounds can directly affect the room-temperature spin polarization. 
\end{abstract}

\maketitle

\section{INTRODUCTION}

To demonstrate highly efficient spin injection and detection in spintronic applications, ferromagnetic Heusler compounds with the chemical formula $X$$_{2}$$Y$$Z$ have been studied,\cite{Groot,Galanakis,Marukame,Inomata,Sakuraba1,Felser,Katsnelson,Tezuka} where $X$ and $Y$ are transition metals and $Z$ is a main group element such as Si and Ge. A schematic diagram of $X$$_{2}$$Y$$Z$ Heusler-compound structures is shown in Fig. 1(a), and four crystal sites  are denoted as A(0,0,0), B($\frac{1}{4}$,$\frac{1}{4}$,$\frac{1}{4}$), C($\frac{1}{2}$,$\frac{1}{2}$,$\frac{1}{2}$), D($\frac{3}{4}$,$\frac{3}{4}$,$\frac{3}{4}$) in Wyckoff coordinates. 

It is well known that Co$_{2}$FeSi (CFS) is one of the Co-based Heusler compounds with the highest Curie temperature and highly spin-polarized density of states (DOS) at the Fermi level.\cite{Wurmehl}  For ideal CFS with an $L$2$_\text{1}$-ordered structure, the (A,C) sites and B sites are occupied with Co and Fe atoms, respectively. Although room-temperature spin-related functionality was shown in magnetic tunnel junctions (MTJs) with CFS electrodes,\cite{Gercsi,Oogane,Inomata2,Tezuka} the spin polarization ($P$) estimated was not so large ($P \le$ 0.5). It has so far been argued that the intentional Co-Fe substitutions between the (A,C) and B sites can affect the decrease in $P$ of CFS electrodes.\cite{Inomata2,Jung,Ksenofontov,Tezuka} Theoretically, electronic band structures were calculated and the effect of the Co-Fe substitutions on $P$ was examined for Co-Fe based Heusler-compound films.\cite{Miyoshi} However, there was no experimental work on the influence of the Co-Fe substitutions on $P$ in the CFS films. 

Recently, in spin light-emitting diodes (LED) with CFS spin injectors,\cite{Ramsteiner,Bruski} highly efficient spin injection into III-V semiconductors was demonstrated. The CFS films were grown on (Al,Ga)As by using low-temperature molecular beam epitaxy (LT-MBE). We also demonstrated relatively high performance for CFS electrodes in lateral spin valves (LSVs),\cite{KimuraCFS} where the CFS films were grown by LT-MBE and highly $L$2$_\text{1}$-ordered structures were also evaluated by conversion electron M\"ossbauer spectroscopy (CEMS) measurements.\cite{Yamada1,Kasahara} Contrary to the previous works with MTJs,\cite{Gercsi,Oogane,Inomata2} relatively large $P \sim$ 0.8 was already obtained at room temperature.\cite{HamayaCFS} We infer that these results originate from the high-quality epitaxial growth with the precise control of the atomic composition using LT-MBE,\cite{Yamada1,Kasahara} and these growth techniques have already been established for other Heusler compounds.\cite{Hamaya1,Yamada2} Hence, if we tune the substitution of Fe for Co using our LT-MBE techniques, the control of $P$ can be demonstrated experimentally and one can clarify the influence of the Co-Fe substitutions on $P$ in CFS. 

In this study, we study the effect of the substitution of Fe for Co on $P$ in low-temperature grown Co$_{3-x}$Fe$_{x}$Si films. Utilizing our LT-MBE techniques, we demonstrate a substitution of Fe for Co occupying A and C sites in Co$_{3-x}$Fe$_{x}$Si films even on group-IV semiconductors. With varying $x$ (1.0 $\le$ $x$ $\le$ 3.0), we clearly observe monotonic change in $P$ at room temperature. This experimental study gives an important knowledge to obtain high-performance Co$_{3-x}$Fe$_{x}$Si Heusler-compound electrodes.

\section{Samples and measurements}
Prior to the fabrication and characterizations of thin-film samples, we briefly explain the crystal structure of general bulk samples by using Fig. 1(a). In the ideal $L$2$_{1}$-type CFS, there are two Co sites (A and C sites), one Fe site (B site), and one Si site (D site), where the A and C sites are magnetically equivalent. With increasing Fe concentration ($x$) from $x$ = 1.0 to 3.0 in Co$_{3-x}$Fe$_{x}$Si, Fe atoms can occupy the A and C sites, leading to the replacement of Co atoms. When all the A and C sites are occupied by Fe atoms, $D$0$_{3}$-type Fe$_{3}$Si (FS) can be obtained.\cite{Niculescu,Hongzhi}
\begin{figure}[t]
\includegraphics[width=8cm]{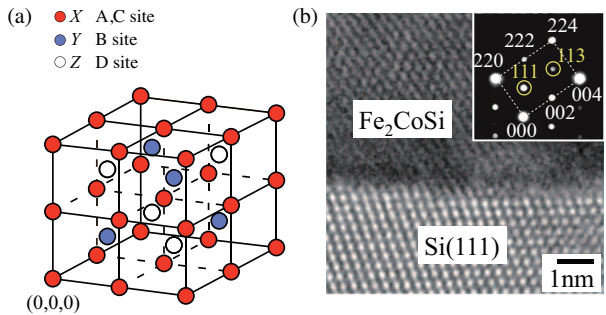}
\caption{(Color online) (a) A schematic diagram of Heusler-compounds with the chemical formula $X$$_{2}$$Y$$Z$. (b) Cross-sectional TEM image of Fe$_{2}$CoSi/Si(111). The inset shows NED patterns for Fe$_{2}$CoSi film near the interface. The zone axis is parallel to the [1$\overline{1}$0] direction.}
\end{figure}  

Co$_{3-x}$Fe$_{x}$Si (1.0 $\le$ $x$ $\le$ 3.0) layers with a thickness of $\sim$25 nm were grown directly on non-doped Si(111) and/or Ge(111) by LT-MBE at 100 $^{\circ}$C, where we co-evaporated Co, Fe and Si using Knudsen cells.\cite{Yamada1,Kasahara} In order to change the ratio of Fe to Co, the growth rates of Co and Fe were tuned by adjusting the cell temperatures. The $x$ value in the films was confirmed by measuring the energy dispersive x-ray spectra. $In$-$situ$ reflection high energy electron diffraction (RHEED) patterns of Co$_{3-x}$Fe$_{x}$Si layers clearly exhibited symmetrical streaks, showing good two-dimensional epitaxial growth. Structural and magnetic properties were characterized by transmission electron microscopy (TEM) images, nanobeam electron diffraction (NED) patterns, magnetization curves, and $^{57}$Fe M\"ossbauer spectra, as shown in our previous works.\cite{Yamada1,Kasahara,Hamaya2,Yamada3} To enhance the detectability of the  $^{57}$Fe M\"ossbauer spectra, we enriched $^{57}$Fe nuclei to 20 \% in the Knudsen cell of the Fe source. For the estimation of the spin polarization at room temperature, the Co$_{3-x}$Fe$_{x}$Si films were patterned into the submicron-sized electrodes by using a conventional electron-beam lithography and an Ar ion milling technique. In order to form the LSVs, 100-nm-thick Cu strips bridging the two Co$_{3-x}$Fe$_{x}$Si electrodes were patterned by a conventional lift-off technique, together with bonding pads. Nonlocal spin-valve (NLSV) measurements were carried out by a conventional current-bias lock-in technique ($\sim 200$ Hz). The detailed fabrication process and measurements are described elsewhere.\cite{HamayaCFS}

\section{Results and Discussion}
Figure 1 shows a high-resolution TEM image of one of the low-temperature grown Co$_{3-x}$Fe$_{x}$Si films, Fe$_{2}$CoSi (FCS, $x$ = 2.0)/Si(111). We see no reaction layers near the interface. Also, lattice image and the NED pattern in the inset of Fig. 1(b) of the FCS layer show single-crystalline characteristics. Superlattice reflections, (111) and (113), originating from the presence of the $D$0$_{3}$- and/or $L$2$_{1}$-structures (solid circles) are clearly identified, consistent with our previous works on CFS ($x =$ 1) \cite{Yamada1,Kasahara} and FS ($x =$ 3).\cite{Hamaya2,Yamada3}  These features imply that the Co atoms occupying the A and C sites are replaced by Fe ones in Co$_{3-x}$Fe$_{x}$Si films with increasing $x$ without forming other disorder structures. For the films grown on Ge(111), we have already confirmed almost identical features. 
\begin{figure}
\includegraphics[width=7.5cm]{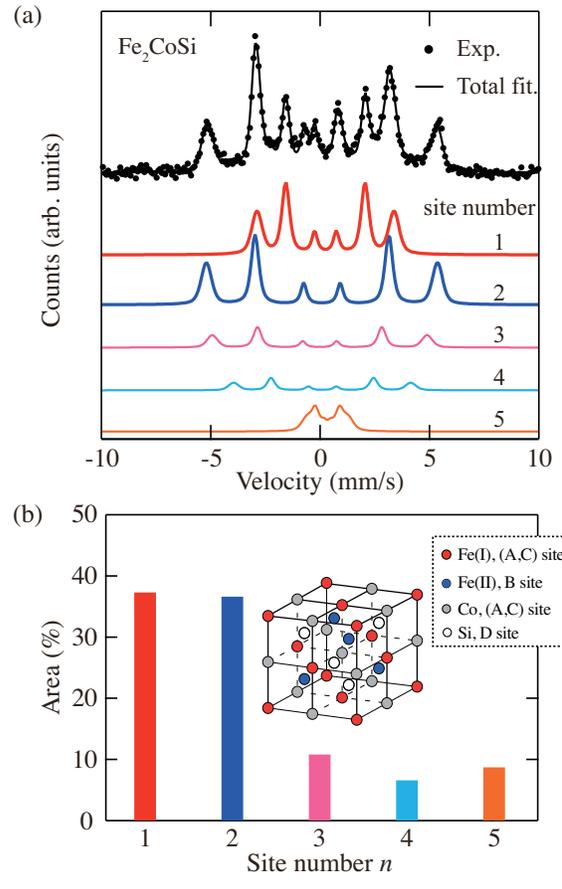}[t]
\caption{(Color online) (a) Room-temperature $^{57}$Fe M\"ossbauer spectrum (solid circles) of an Fe$_{2}$CoSi film, together with five fitting subspectra as denoted in the text. (b) Fitted area versus site number for Fe$_{2}$CoSi film, estimated by the CEMS measurement and best fitting. The inset shows schematic diagram of an ordered Fe$_{2}$CoSi structure. There are two types of magnetically distinct Fe sites, i.e., Fe(I), occupying A or C sites coordinated with four Fe atoms and Si atoms, and Fe(II), occupying the B site surrounded by eight Fe atoms or Co ones.}
\end{figure}  

In order to verify the atomic substitution between Co and Fe, we examined the magnetic environments around the Fe sites for a FCS film ($x$ = 2.0) by $^{57}$Fe M\"ossbauer spectroscopy. One of the crystal structures of an ordered FCS is illustrated in the inset of Fig. 2(b). There are two types of magnetically distinct Fe sites, i.e., Fe(I), occupying A or C site coordinated with four Fe atoms and four Si atoms, and Fe(II), occupying the B site surrounded by eight Fe or Co atoms. In a previous work of bulk Co$_{3-x}$Fe$_{x}$Si samples reported by Jung {\it et al}.,\cite{Jung} the two sites with two different hyperfine magnetic fields were completely distinguished by the $^{57}$Fe M\"ossbauer spectroscopy. Thus, we can probe the local structures of the FCS film by detecting the two different surroundings of Fe atoms. The $^{57}$Fe M\"ossbauer spectroscopy for the FCS epilayer grown is presented in Fig. 2(a). An evident sextet pattern, typical for magnetically ordered systems, is obtained (black solid circles). The spectrum can be successfully fitted with five magnetic environments defined as site number $n$ from 1 to 5, where the five fitting curves are also presented in  Fig. 2(a). Note that the quadrupole shift was fixed to zero due to the cubic symmetry of the local Fe environment and the intensity ratio of the magnetically split sextet was also fixed to 3$:$4$:$1$:$1$:$4$:$3, since the magnetic moment was along the film plane, reflecting the magnetic shape anisotropy. The large two contributions, denoted by the site 1 and 2 with hyperfine magnetic fields of 19.4 and 32.8 T, respectively, are the same as the two sites, Fe(I) and Fe(II), in the bulk samples in Ref.\cite{Jung}. On the other hand, the other three subspectra with smaller intensity can be recognized as disordered sites. 
\begin{figure}[t]
\includegraphics[width=7.5cm]{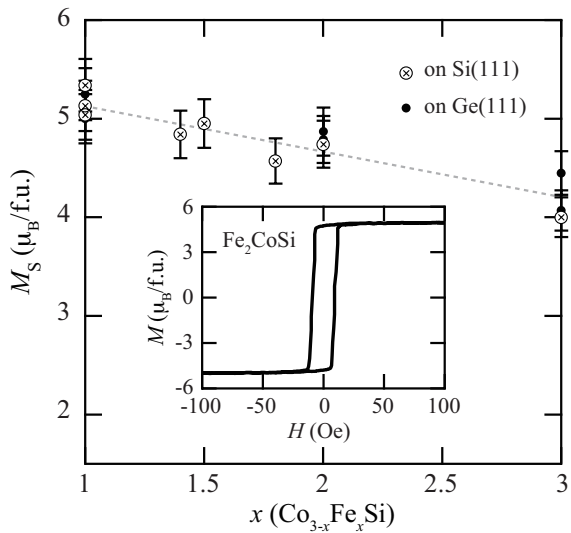}
\caption{(a) $M$$_{\rm S}$ as a function of $x$ for Co$_{3-x}$Fe$_{x}$Si films on Si(111) and/or Ge(111), measured at 300 K. The inset shows the $M$-$H$ curve for Fe$_{2}$CoSi, measured at 300 K.}
\end{figure} 

The percentage of the fitted area versus site number for the FCS film is also presented in Fig. 2(b). If we obtain the perfectly substituted structure in FCS ($x$ = 2.0), there are Fe(I) and Fe(II) sites evenly, i.e., site 1 $:$ site 2 $=$ 50 (\%) $:$ 50 (\%). In actual our FCS epilayer, site 1 $:$ site 2 $=$ 37.3 (\%) $:$ 36.6 (\%) , indicating that the local degree of the structural ordering exceeds 70 \% even for the low-temperature grown FCS film. We note that the ratio of site 1 to site 2 is nearly one to one. Since the hyperfine magnetic fields for the site 3 and site 4 were 30.5 T and 25.1 T, respectively, the site 3 and 4 might be very similar to Fe(II) and Fe(I), respectively. Such deviations of the hyperfine magnetic fields from the ideal values can be caused by the site occupation of surplus Fe and Si atoms with (A,C) sites.\cite{Hamaya2,Yamada3} In other words, although the presence of about 15 \% sites (sites 3 and 4) was also arising from the Fe(I)- and Fe(II)-like structures, local off-stoichiometry led to such disordered structures. Finally, the site 5 with less than 15 \% was regarded as the almost nonmagnetic site with a small hyperfine magnetic field of less than 5 T, which may originate from the FCS/Ge interface.\cite{Yamada3}

In our previous works for FS ($x =$ 3),\cite{Hamaya2,Yamada3} we have obtained relatively high local degree of the structural ordering ($\sim$ 70 \%) for FS ($x =$ 3). Also, we have confirmed that the local degree of the structural ordering exceeds 70 \% is achieved and the ratio of site 1 to site 2 is nearly one to one in Fig. 2. Thus, even if we conduct the change in $x$ in LT-MBE conditions, we can achieve the atomic substitutions without decreasing the structural ordering in the Heusler-compound structure. Considering the fact that there was no site occupation of Fe atoms with D sites for CFS ($x =$ 1),\cite{Yamada1,Kasahara} we can eventually judge that the Fe atoms are substituted for the Co ones occupying with A and C sites, following the site preference selectivity described for a bulk Co$_{3-x}$Fe$_{x}$Si system.\cite{Niculescu} 

Taking these structural characterizations into account, we further examined magnetic properties with increasing $x$ in Co$_{3-x}$Fe$_{x}$Si films grown on Si(111) and/or Ge(111). The inset of Fig. 3 shows representative field-dependent magnetization ($M-H$) curve at 300 K for the FCS film, where the applied field direction is parallel to the magnetic easy axis in the film plane. A clear ferromagnetic hysteretic curve is observed and the coercive force is very small with a value of $\sim$9 Oe, which is consistent with typical characteristics of Heusler compounds.\cite{Groot,Galanakis,Marukame,Inomata,Sakuraba1,Felser,Katsnelson,Sagar} The estimated saturation magnetization ($M$$_{\rm S}$) reaches more than $\sim$80 \% of bulk $M$$_{\rm S}$.\cite{Niculescu,Hongzhi} These features were seen for all the Co$_{3-x}$Fe$_{x}$Si films (1.0 $\le$ $x$ $\le$ 3.0), implying that the crystal quality of the Co$_{3-x}$Fe$_{x}$Si films are very good. Then, we summarize $M$$_{\rm S}$ versus $x$ for the Co$_{3-x}$Fe$_{x}$Si films in the main panel of Fig. 3, where $M$$_{\rm S}$ values of the Co$_{3-x}$Fe$_{x}$Si films grown on Si and Ge are plotted. We can find that $M$$_{\rm S}$ is monotonically tuned with varying $x$. This tendency is in good agreement with that of bulk samples in a previous work,\cite{Niculescu} in which this behavior in $M$$_{\rm S}$ versus $x$ can be explained by the change in the local magnetic moment of Fe atoms due to the substitution of Fe for Co occupying A and C sites.\cite{Niculescu} Therefore, we conclude that the obtained monotonic change in $M$$_{\rm S}$ versus $x$ results from the experimentally verified atomic substitution in Co$_{3-x}$Fe$_{x}$Si films (1.0 $\le$ $x$ $\le$ 3.0).

Also, the electronic structures are determined predominantly by the chemical surroundings of magnetic atoms such as the number and type of the nearest neighbors at given site.\cite{Inomata2,Nakatani,Hongzhi,Hamaya1,Miyoshi} Since the magnetic properties in Co$_{3-x}$Fe$_{x}$Si films on Si or Ge were controlled by the Co-Fe atomic substitution, the control of $P$ can also be expected because of the systematic change in the DOS at the Fermi level.\cite{Inomata2,Jung,Ksenofontov} To confirm this effect, we finally evaluated the room-temperature $P$ for Co$_{3-x}$Fe$_{x}$Si films by using NLSV measurements and the analysis based on the one-dimensional spin diffusion model.\cite{KimuraCFS,HamayaCFS,Oki} Since we have so far evaluated CFS ($x =$ 1.0) and FS ($x =$ 3.0),\cite{HamayaCFS} we hereafter focus on Fe$_{1.5}$Co$_{1.5}$Si ($x =$ 1.5) and FCS ($x =$ 2.0). A scanning electron micrograph (SEM) of a representative FCS/Cu LSV is shown in Fig. 4(a). The center-to-center separation between FCS injector and detector ($d$) is 400 nm. Figure 4(b) shows a nonlocal magnetresistance curve of a FCS/Cu LSV, measured at room temperature. We observe clear NLSV signal depending on the magnetization configuration between the spin injector and detector. It should be noted that a relatively large spin signal ($\Delta R_{\rm S}$) of $\sim$1.3 m$\Omega$ is seen, which is roughly five times larger than that for Py/Cu LSVs of the same size and low-resistance ohmic interfaces.\cite{KimuraCFS}
\begin{figure}[t]
\includegraphics[width=8.5cm]{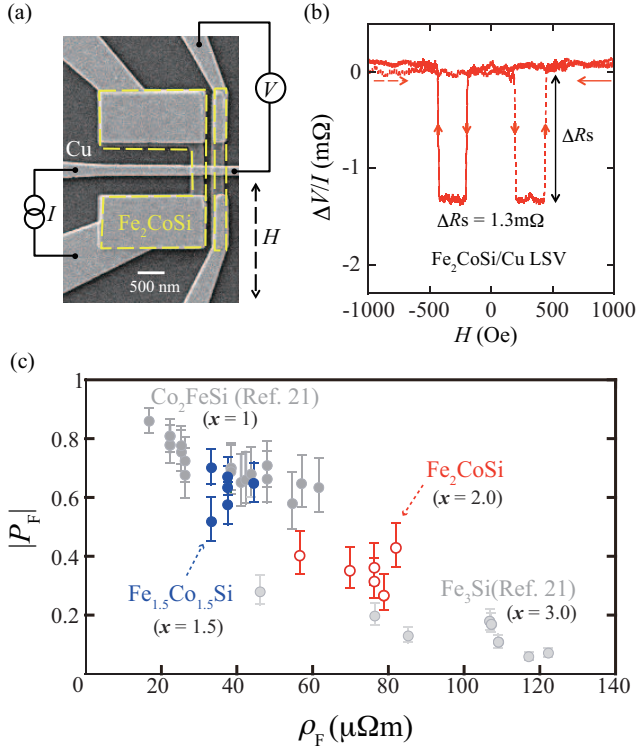}
\caption{(Color online) (a) Scanning electron microscope image of fabricated Fe$_{2}$CoSi/Cu LSV. (b) Room-temperature NLSV signal for Fe$_{2}$CoSi/Cu LSV. (c) Room-temperature $P_{\rm F}$ as a function of $\rho$$_\text{F}$ for Co$_{3-x}$Fe$_{x}$Si/Cu LSVs. For $x =$ 1.0 and 3.0, we replotted the data presented in Ref. \cite{HamayaCFS}. The $\lambda$$_\text{F}$ values used are 5 $\pm$ 1 nm ($x =$ 3.0), 4 $\pm$ 1 nm ($x =$ 2.0 and 1.5), and 3 $\pm$ 1 nm ($x =$ 1.0).}
\end{figure} 

In our previous works,\cite{KimuraCFS,HamayaCFS,Oki} we simplified the spin signals on the basis of a one-dimensional spin diffusion model:\cite{VFmodel,Takahashi,Kimura}
\begin{equation}
\Delta R_{\rm S} A \approx 
\frac{ \left( \frac{ P_F}{(1-P_F^2)} \rho_{\rm F} \lambda_{\rm F} 
+ \frac{P_I}{(1-P_I^2)} RA_{\rm F/N} \right)^2 }
{\rho_{\rm N} \lambda_{\rm N} \sinh \left( {d}/{\lambda_{\rm N}} \right)},
\end{equation}
where $P_{\rm F}$, $P_{\rm I}$, $\rho$$_\text{F}$, and $\lambda$$_\text{F}$ are the bulk spin polarization, interface spin polarization, resistivity, and spin diffusion length of the ferromagnetic electrode, respectively. $\rho$$_\text{N}$ and $\lambda$$_\text{N}$ are those for the nonmagnetic wire, and $d$ is the separation distance between the injector and detector ($d =$ 400 nm). We determined $\rho$$_\text{F}$ by the four-terminal resistance measurements. Also, we know that $\rho$$_\text{N}$ and $\lambda$$_\text{N}$ are 2.5 $\mu$$\Omega$cm and 500 nm, respectively.\cite{KimuraCFS,HamayaCFS} $RA$$_\text{F/N}$ is the resistance of the interface between the ferromagnet and nonmagnet, but this term can be ignored since we have already confirmed that $RA_{\rm F/N}$ $<$ 0.1 f$\Omega$m$^{2}$. Note that the area parameter $A$ is defined as  ($S_{\rm inj}$$S_{\rm det}$/$S_{\rm N}$), where $S_{\rm inj}$, $S_{\rm det}$, and $S_{\rm N}$ are the areas of the junctions in the spin injector and spin detector, and the cross section of the nonmagnetic strip, respectively as described in our previous works.\cite{KimuraCFS,HamayaCFS,Oki} The $A$ values for the  Co$_{3-x}$Fe$_{x}$Si/Cu junctions were directly measured by SEM observations. Hence, using Eq. (1), we can obtain $P$$_\text{F}$ by making an assumption about the value of $\lambda$$_\text{F}$. In this study, we used the $\lambda$$_\text{F}$ value of 4 $\pm$ 1 nm for $x$ = 2.0 (FCS) and 1.5 (Fe$_{1.5}$Co$_{1.5}$Si) considering our previous works.\cite{KimuraCFS,HamayaCFS} The resulting $P$$_\text{F}$ values estimated as a function of $\rho$$_\text{F}$ for the Co$_{3-x}$Fe$_{x}$Si films ($x$ = 1.0, 1.5, 2.0, and 3.0) are shown in Fig. 4(c). There is an almost monotonic change in the magnitude of $P$$_\text{F}$ with varying $\rho$$_\text{F}$, which implies that the structural ordering crucially affects the $P_{\rm F}$ values for Co$_{3-x}$Fe$_{x}$Si films, consistent with general interpretations in Heusler compounds.\cite{HamayaCFS,Ploog,Schneider,Blum,Ramsteiner,Bruski} 

We note that the magnitude of $P$$_\text{F}$ values for the Co$_{3-x}$Fe$_{x}$Si/Cu LSVs are also changed monotonically with varying $x$. Since the resistivity shown in horizontal axis depended on the microscopic structural ordering in the Heusler wires, the scattering of the horizontal axis could not be avoided.\cite{HamayaCFS} Also, the sign of $P$$_\text{F}$ depending on the structural ordering should be considered.\cite{Bruski,Oki2} However, Fig. 4(c) apparently reveals that the room-temperature $P$$_\text{F}$ can be tuned by systematically introducing Co-Fe substitution in Co$_{3-x}$Fe$_{x}$Si films even on group-IV semiconductor platforms. 

\section{Summary}
We have studied the effect of the Co-Fe substitution in Co$_{3-x}$Fe$_{x}$Si on $P$ by using LT-MBE techniques and nonlocal spin valve studies. Even for the low-temperature grown Heusler-compound films, the Co-Fe atomic substitution at A and C sites can be confirmed by the conversion electron M\"ossbauer spectroscopy measurements. Furthermore, the magnetic moment is systematically changed with the Co-Fe substitutions. The room-temperature $P$ was monotonically tuned by systematically changing $x$. This study experimentally verified that the Co-Fe substitution in Co$_{3-x}$Fe$_{x}$Si Heusler compounds can sensitively affect the room-temperature spin polarization. 

\begin{acknowledgements}
The authors would like to thank Prof. T. Kimura of Kyushu University for their useful discussions. 
This work was partially supported by CREST-JST, STARC, SCOPE-MIC, TEPCO Memorial Foundation, and Grant-in-Aid for challenging Exploratory Research from JSPS.  The M\"ossbauer measurements were supported by the Nanotechnology Network Platform of MEXT, Japan. 
\end{acknowledgements}

\noindent{{\bf References}}


\begin{thebibliography}{11}
\bibitem{Groot}
R. A. de Groot, F. M. Mueller, P. G. van Engen, and K. H. J. Buschow, Phys. Rev. Lett. {\bf 50}, 2024 (1983).
\bibitem{Galanakis}
I. Galanakis, P. H. Dederichs, and N. Papanikolaou, Phys. Rev. B {\bf 66}, 174429 (2002). 
\bibitem{Inomata}
K. Inomata, S. Okamura, R. Goto, and N. Tezuka, Jpn. J. Appl. Phys. {\bf 42}, L419 (2003).
\bibitem{Marukame}
T. Marukame, T. Kasahara, K. Matsuda, T. Uemura, and M. Yamamoto, Jpn. J. Appl. Phys., Part 2 {\bf 44}, L521 (2005). 
\bibitem{Sakuraba1}
Y. Sakuraba, M. Hattori, M. Oogane, Y. Ando, H. Kato, A. Sakuma, T. Miyazaki, and H. Kubota, Appl. Phys. Lett. {\bf 88}, 192508 (2006). 
\bibitem{Felser}
G. H. Fecher and C. Felser, J. Phys. D: Appl. Phys. {\bf 40}, 1582 (2007).
\bibitem{Katsnelson}
M. I. Katsnelson, V. Yu. Irkhin, L. Chioncel, A. I. Lichtenstein, R. A. de Groot, Rev. Mod. Phys.  {\bf 80}, 315 (2008). 
\bibitem{Tezuka}
N. Tezuka, N. Ikeda, F. Mitsuhashi, and S. Sugimoto, Appl. Phys. Lett. {\bf 94}, 162504 (2009). 

\bibitem{Wurmehl}
S. Wurmehl, G. H. Fecher, H. C. Kandpal, V. Ksenofontov, C. Felser, H.-J. Lin, and J. Morais, Phys. Rev. B {\bf 72}, 184434 (2005). 
\bibitem{Gercsi}
Z. Gercsi, A. Rajanikanth, Y. K. Takahashi, K. Hono, M. Kikuchi, N. Tezuka, and K. Inomata, Appl. Phys. Lett. {\bf 89}, 082512 (2006).
\bibitem{Inomata2}
K. Inomata, N. Ikeda, N. Tezuka, R. Goto, S. Sugimoto, M. Wojcik, and E. Jedryka, Sci. Technol. Adv. Mater. {\bf 9}, 014101 (2008).
\bibitem{Oogane}
M. Oogane, M. Shinano, Y. Sakuraba, and Y. Ando, J. Appl. Phys. {\bf 105}, 07C903 (2009).
\bibitem{Jung}
V. Jung, B. Balke, G. H. Fecher, and C. Felser, Appl. Phys. Lett. {\bf 93}, 042507 (2008).
\bibitem{Ksenofontov}
V. Ksenofontov, M. W\'{o}jcik, S. Wurmehl, H. Schneider, B. Balke, G. Jakob, and C. Felser, J. Appl. Phys. {\bf 107}, 09B106 (2010).
\bibitem{Miyoshi}
H. Mori, Y. Odahara, D. Shigyo, T. Yoshitake, E. Miyoshi, Thin Solid Films {\bf 520}, 4979 (2012).

\bibitem{Ramsteiner}
M. Ramsteiner, O. Brandt, T. Flissikowski, H. T. Grahn, M. Hashimoto, J. Herfort, and H. Kostial, Phys. Rev. B {\bf 78}, 121303(R) (2008).
\bibitem{Bruski}
P. Bruski, S. C. Erwin, M. Ramsteiner, O. Brandt, K.-J. Friedland, R. Farshchi, J. Herfort, and H. Riechert, Phys. Rev. B {\bf 83}, 140409(R) (2011).

 
\bibitem{KimuraCFS}
T. Kimura, N. Hashimoto, S. Yamada, M. Miyao, K. Hamaya, NPG asia materials {\bf 4}, e9 (2012).
\bibitem{Yamada1}
S. Yamada, K. Hamaya, K. Yamamoto, T. Murakami, K. Mibu, and M. Miyao, Appl. Phys. Lett. {\bf 96}, 082511 (2010). 
\bibitem{Kasahara}
K. Kasahara, K. Yamamoto, S. Yamada, T. Murakami, K. Hamaya, K. Mibu, and M. Miyao, J. Appl. Phys. {\bf 107}, 09B105 (2010).
\bibitem{HamayaCFS}
K. Hamaya, N. Hashimoto, S. Oki, S. Yamada, M. Miyao, and T. Kimura, Phys. Rev. B {\bf 85}, 100404(R) (2012).

\bibitem{Hamaya1}
K. Hamaya, H. Itoh, O. Nakatsuka, K. Ueda, K. Yamamoto, M. Itakura, T. Taniyama, T. Ono, and M. Miyao, Phys. Rev. Lett. {\bf 102}, 137204 (2009).
\bibitem{Yamada2}
S. Yamada, K. Hamaya, T. Murakami, B. Varaprasad, Y. K. Takahashi, A. Rajanikanth, K. Hono, and M. Miyao, J. Appl. Phys. {\bf 109}, 07B113 (2011).

\bibitem{Niculescu}
V. Niculescu, J. I. Budnick, W. A. Hines, K. Raj, S. Pickart, and S. Skalski, Phys. Rev. B {\bf 19}, 452 (1979).
\bibitem{Hongzhi}
L. Hongzhi, Z. Zhiyong, M. Li, X. Shifeng, L. Heyan, Q. Jingping, L. Yangxian, and W. Guangheng, J. Phys. D: Appl. Phys. {\bf 40}, 7121 (2007).


\bibitem{Hamaya2}
K. Hamaya, T. Murakami, S. Yamada, K. Mibu, and M. Miyao, Phys. Rev. B {\bf 83}, 144411 (2011).
\bibitem{Yamada3}
S. Yamada, J. Sagar, S. Honda, L. Lari, G. Takemoto, H. Itoh, A. Hirohata, K. Mibu, M. Miyao, and K. Hamaya, Phys. Rev. B {\bf 86}, 174406 (2012).
\bibitem{Sagar}
J. Sagar, H. Sukegawa, L. Lair, V. K. Lazarov, S. Mitani, K. O'Grady, and A. Hirohata, Appl. Phys. Lett. {\bf 101}, 102410 (2012). 

\bibitem{Nakatani}
T. M. Nakatani, A. Rajanikanth, Z. Gercsi, Y. K. Takahashi, K. Inomata, and K. Hono, J. Appl. Phys. {\bf 102}, 033916 (2007).

\bibitem{Oki}
S. Oki, S. Yamada, N. Hashimoto, M. Miyao, T. Kimura, and K. Hamaya, Appl. Phys. Exp. {\bf 5}, 063004 (2012).
\bibitem{VFmodel}
T. Valet, and A. Fert, Phys. Rev. B {\bf 48}, 7099 (1993).
\bibitem{Takahashi}
S. Takahashi and S. Maekawa, Phys. Rev. B {\bf 67}, 052409 (2003).
\bibitem{Kimura}
T. Kimura, J. Hamrle, and Y. Otani, Phys. Rev. B {\bf 72}, 014461 (2005); T. Kimura, and Y. Otani, J. Phys. Condens. Matter. {\bf 19}, 165216 (2007).
\bibitem{Ploog}
J. Herfort, H.-P. Sch\"onherr, K.-J. Friedland, and K. H. Ploog, J. Vac. Sci. Technol. B {\bf 22}, 2073 (2004); M. Hashimoto, J. Herfort, H.-P. Sch\"onherr, and K. H. Ploog, J. Appl. Phys. {\bf 98}, 104902 (2005); J. Herfort, B. Jenichen, V. Kaganer, A. Trampert, H.-P. Sch\"onherr, and K. H. Ploog, Physica E {\bf 32}, 371 (2006).
\bibitem{Schneider}
H. Schneider, G. Jakob, M. Kallmayer, H. J. Elmers, M. Cinchetti, B. Balke, S. Wurmehl, C. Felser, M. Aeschlimann, and H. Adrian, Phys. Rev. B {\bf 74}, 174426 (2006).
\bibitem{Blum}
C. G. F. Blum, C. A. Jenkins, J. Barth, C. Felser, S. Wurmehl, G. Friemel, C. Hess, G. Behr, B. B\"uchner, A. Reller, S. Riegg, S. G. Ebbinghaus, T. Ellis, P. J. Jacobs, J. T. Kohlhepp, and H. J. M. Swagten, Appl. Phys. Lett. {\bf 95}, 161903 (2009).

\bibitem{Oki2}
S. Oki, K. Masaki, N. Hashimoto, S. Yamada, M. Miyata, M. Miyao, T. Kimura, and K. Hamaya, Phys. Rev. B {\bf 74}, 174412 (2012).


%



\end{thebibliography}
\end{document}